\documentclass{article}

\PassOptionsToPackage{numbers, sort&compress}{natbib}

\usepackage[final]{neurips_2025}

\def\jcap{JCAP}
\def\mnras{MNRAS}
\def\apj{ApJ}
\def\prd{PRD}
\def\baas{BAAS}

\usepackage[utf8]{inputenc} % allow utf-8 input
\usepackage[T1]{fontenc}    % use 8-bit T1 fonts
\usepackage{hyperref}       % hyperlinks
\usepackage{url}            % simple URL typesetting
\usepackage{booktabs}       % professional-quality tables
\usepackage{amsfonts}       % blackboard math symbols
\usepackage{nicefrac}       % compact symbols for 1/2, etc.
\usepackage{microtype}      % microtypography
\usepackage{xcolor}         % colors
\usepackage{bm}% bold math
\usepackage{mathtools}      
\usepackage{multirow}
\usepackage{physics}     
\usepackage{amsmath}
\usepackage{amssymb}
\usepackage{graphicx}
\usepackage{enumerate}
\usepackage{enumitem}

\definecolor{hypercolor}{RGB}{174, 60, 60} 
\hypersetup{
  linkcolor  = hypercolor,
  citecolor  = hypercolor,
  urlcolor   = hypercolor,
  colorlinks = true
}

\title{Vision Transformers for Cosmological Fields: Application to Weak Lensing Mass Maps}

\author{%
  Jash Kakadia \\
    Department of Physics and Astronomy\\
    University of Pennsylvania\\
    Philadelphia, PA 19104\\
  \texttt{jkakadia@seas.upenn.edu} \\
  \And
  Shubh Agrawal \\
  Department of Physics and Astronomy\\
    University of Pennsylvania\\
    Philadelphia, PA 19104\\
  \texttt{shubh@sas.upenn.edu} \\
  \AND
  Kunhao Zhong \\
  Department of Physics and Astronomy\\
    University of Pennsylvania\\
    Philadelphia, PA 19104\\
  \texttt{kunhaoz@sas.upenn.edu} \\
  \And
  Bhuvnesh Jain \\
  Department of Physics and Astronomy\\
    University of Pennsylvania\\
    Philadelphia, PA 19104\\
  \texttt{bjain@physics.upenn.edu}
}

\makeatletter
\def\@trackname{\@neuripsordinal\ Conference on Neural Information Processing Systems (NeurIPS \@neuripsyear).}
\makeatother

\begin{document}

\maketitle

\begin{abstract}
Weak gravitational lensing is a powerful probe of the universe’s growth history. While traditional two-point statistics capture only the Gaussian features of the convergence field, deep learning methods such as convolutional neural networks (CNNs) have shown promise in extracting non-Gaussian information from small-scale, nonlinear structures. In this work, we evaluate the effectiveness of attention-based architectures, including variants of vision transformers (ViTs) and shifted window (Swin) transformers, in constraining the cosmological parameters $\Omega_m$ and $S_8$ from weak lensing mass maps. Using a simulation-based inference (SBI) framework, we compare transformer-based methods to CNNs. We also examine performance scaling with the number of available $N$-body simulations, highlighting the importance of pre-training for transformer architectures. We find that the Swin transformer performs significantly better than vanilla ViTs, especially with limited training data. Despite their higher representational capacity, the Figure of Merit for cosmology achieved by transformers is comparable to that of CNNs under realistic noise conditions.
\end{abstract}

\section{Introduction}\label{sec:intro}

Spatial fluctuations in matter density encode information about the cosmological composition and evolution of the universe. Weak gravitational lensing is a key probe of structure formation, as it does not rely on modeling biased tracers (galaxies). Stage III experiments, such as the Dark Energy Survey \cite{amon_dark_2022, secco_dark_2022}, Hyper Suprime-Cam Survey \cite{miyatake2023hyper}, and Kilo-Degree Survey \cite{hildebrandt_kids-450_2017, heymans_kids-1000_2021}, have placed leading constraints on the matter density $\Omega_\mathrm{m}$ and fluctuation amplitude $\sigma_8$. However, the persistent tension between early- and late-time $\sigma_8$ measurements suggests possible new physics beyond $\Lambda$CDM model. This motivates the development of alternative statistics that go beyond the traditional two-point function, which will be even more important for upcoming surveys such as Euclid \cite{laureijs2011euclid}, LSST \cite{LSST:2008ijt}, and the Roman Space Telescope \cite{Dore:2019pld}.

On large scales, weak lensing maps are nearly Gaussian and well described by two-point statistics \cite{hildebrandt_kids-450_2017, Troxel2018}. On small scales, however, nonlinear gravitational evolution induces non-Gaussian features, as overdensities collapse into halos and galaxies \cite{kilbinger_cosmology_2015}. A variety of statistics have been developed to capture this information, including higher-order correlations \cite{zaldarriaga_higher_2003, takada_three-point_2003}, peak counts \cite{kratochvil_probing_2010, dietrich_cosmology_2010}, Minkowski functionals \cite{parroni_going_2020}, scattering transforms \cite{cheng_new_2020}, and deep learning approaches \cite{gupta_non-gaussian_2018, fluri_cosmological_2018, lu_simultaneously_2022, Zhong:2024qpf}. These have been applied to both simulations and real data \cite{petri_emulating_2015, gatti_dark_2020, valogiannis_going_2022, lu_cosmological_2023}.

Convolutional neural networks (CNNs) have been widely used in weak lensing to extract non-Gaussian information by mapping convergence fields to cosmological parameters. In this work, we explore attention-based models (Transformers). Transformers were originally developed for natural language processing \cite{vaswani2017attention} and later adapted for images \cite{dosovitskiy_image_2021, carion_end--end_2020}. They process inputs in parallel using self-attention and have now found astronomical applications in galaxy morphology classification \cite{lin_galaxy_2022}, strong gravitational lens parameter estimation \cite{huang_strong_2022}, target detection in multi-band photometric surveys \cite{jia_deep_2023}, and cosmology from 3D dark matter halo lightcones \cite{hwang_universe_2023}.

ViTs are especially attractive for weak lensing applications. They capture global context directly through attention, without requiring a hierarchical stack of local filters as in CNNs. This enables more efficient learning of long-range dependencies compared to CNNs, which learn such dependencies only in deeper layers. ViTs also have weaker inductive biases, which allows for more flexible representations but increases the need for training data and careful regularization. A key advantage is interpretability: attention weights can be analyzed using tools such as attention rollout, attention flow \cite{abnar_quantifying_2020}, and relevance propagation \cite{chefer_transformer_2021}.

We present the first application of Vision Transformers—and more generally, self-attention-based deep learning—for predicting cosmological parameters from simulated weak lensing maps. Alongside three configurations of the original ViT \cite{dosovitskiy_image_2021}, we study shifted window transformers (Swin Transformers) \cite{2021arXiv210314030L}. These use local self-attention within hierarchical windows, offering improved efficiency and better scaling to high-resolution inputs.

Our manuscript is organized as follows. In Section~\ref{sec:methodology}, we introduce simulation-based inference, how vision models can be used to constrain cosmology from the field level, the simulations we use, and how we train the models. In Section~\ref{sec:models}, we review the key differences of attention-based models. We present the results and conclude in Section~\ref{sec:results}.
\section{Methodology}\label{sec:methodology}

Simulation-based inference (SBI), also known as likelihood-free inference (LFI), enables Bayesian parameter inference directly from simulated observables. It is especially useful when the likelihood is intractable, but forward simulations are feasible, like the case of higher-order statistics. Neural posterior estimation (NPE), or density-estimation LFI (DELFI), adopted this approach by training neural density estimators to learn the relationship between parameters and data. These methods can achieve posterior inference using orders of magnitude fewer simulations than traditional LFI approaches \cite{bonassi_bayesian_2011, fan2012approximate, papamakarios2018fast, lueckmann2019likelihoodfree}. DELFI has been applied to supernova analysis \cite{alsing_massive_2018}, high-redshift Ly$\alpha$ forest modeling, tomographic weak lensing \cite{alsing_fast_2019}, and CMB parameter inference \cite{Lemos_2023}. For an overview, see \cite{2021arXiv210104653L}.

We frame our inference as a conditional density estimation task. Given simulation parameters $\theta$ and compressed data vectors $d$, we train an ensemble of neural density estimators (NDEs) to approximate $p(d|\theta)$, by auto-regressively reducing the Kullback–Leibler distance. Assuming a flat prior matching the simulation distribution, this yields posteriors from deep learning models. Our implementation uses the public \texttt{sbi} package. The ensemble consists of three Masked Autoregressive Flows (MAFs) and three Gaussian Mixture Density Networks (MDNs), all trained separately. The MAFs have 2, 3, and 4 MADE layers with 50, 50, and 25 transforms per layer, respectively. The MDNs have 2, 3, and 4 Gaussian components, using 50, 50, and 25 neurons per layer.

In the limit of abundant simulations, all NDEs should yield consistent posteriors. To reduce outlier effects, we impose a selection criterion: we discard any NDE whose $1\sigma$ constraint on $S_8$ deviates by more than 5\% from the ensemble median. The remaining NDEs' samples are concatenated to form the final posterior. We validate the inference using the empirical coverage test \cite{2017arXiv170604599G, 2021arXiv211006581H, 2023PMLR..20219256L}, implemented via the public \texttt{TARP} package. This test assesses whether the inferred posteriors are well-calibrated, i.e. whether they are too conservative or too narrow—by comparing against ground-truth parameter values across repeated trials.

We use weak lensing convergence maps generated from the \texttt{DarkGridV1} $N$-body simulation suite \cite{Zuercher2021, Zuercher2021b}. The convergence maps are constructed as weighted projections of 3D overdensity fields, following procedures similar to \cite{sourceclustering, Gatti2023}. We adopt the DES-Y3 redshift distribution for four tomographic bins, but rescale the galaxy number density and survey area to match LSST Year 1. While real LSST-Y1 data will have different galaxy selections and lensing kernels, our conclusions are not sensitive to this, as the dominant factor remains the noise level. The simulations vary only $\Omega_\mathrm{m}$ and $\sigma_8$, while holding other cosmological parameters fixed and omitting systematics such as photometric redshift uncertainty. These simplifications are acceptable for our purpose of comparing different architectures. For testing purposes only, we also made a shape-noise free simulation set with only one channel to test the model difference with maximum signal-to-noise ratio.

For each configuration, we use an 80\%:10\%:10\% for our training, validation, and testing split. In total there are 13680 maps in the dataset, and each input is a $512\times512$ independent flattened patch of a convergence map, with either one or four channels. We train each network independently using the AdamW optimizer \cite{loshchilov2019decoupledweightdecayregularization}, with an early stopping patience of 30 epochs. A  L2 error metric (root-mean-squared, RMSE) is used, computed over the validation dataset. Batch size is 32, except for the much larger networks \texttt{ViT-L} and \texttt{ViT-H} (each $>300$M parameters), where we set to 24 \textit{(16)} and 12 \textit{(8)} respectively for the one-channel \textit{(four-channel)} dataset, due to GPU memory constraints (we use Nvidia A100s with 80 GBs of memory). We use the PyTorch ReduceLROnPlateau scheduler with a 10-epoch patience and reduction rate of 0.3. For all convolutional networks and vision transformers, the initial learning rate was set to $10^{-3}$ and $10^{-5}$ respectively. Other values were set to PyTorch defaults.

Vision Transformers are known to require large training sets \cite{dosovitskiy_image_2021}. To address this, we explore pre-training on synthetic data to improve performance. We generate pre-training data following \cite{Zhong:2024wdk}, which uses an analytic model to rapidly produce maps with similar statistics to weak lensing fields. Our synthetic dataset includes 13680  maps, on which we pre-train the models in a fashion identical to the above. We then use the same method to train the pre-trained models on the real dataset.

\begin{figure}\label{fig:s8_residuals}
    \centering
    \includegraphics[width=0.49\linewidth]{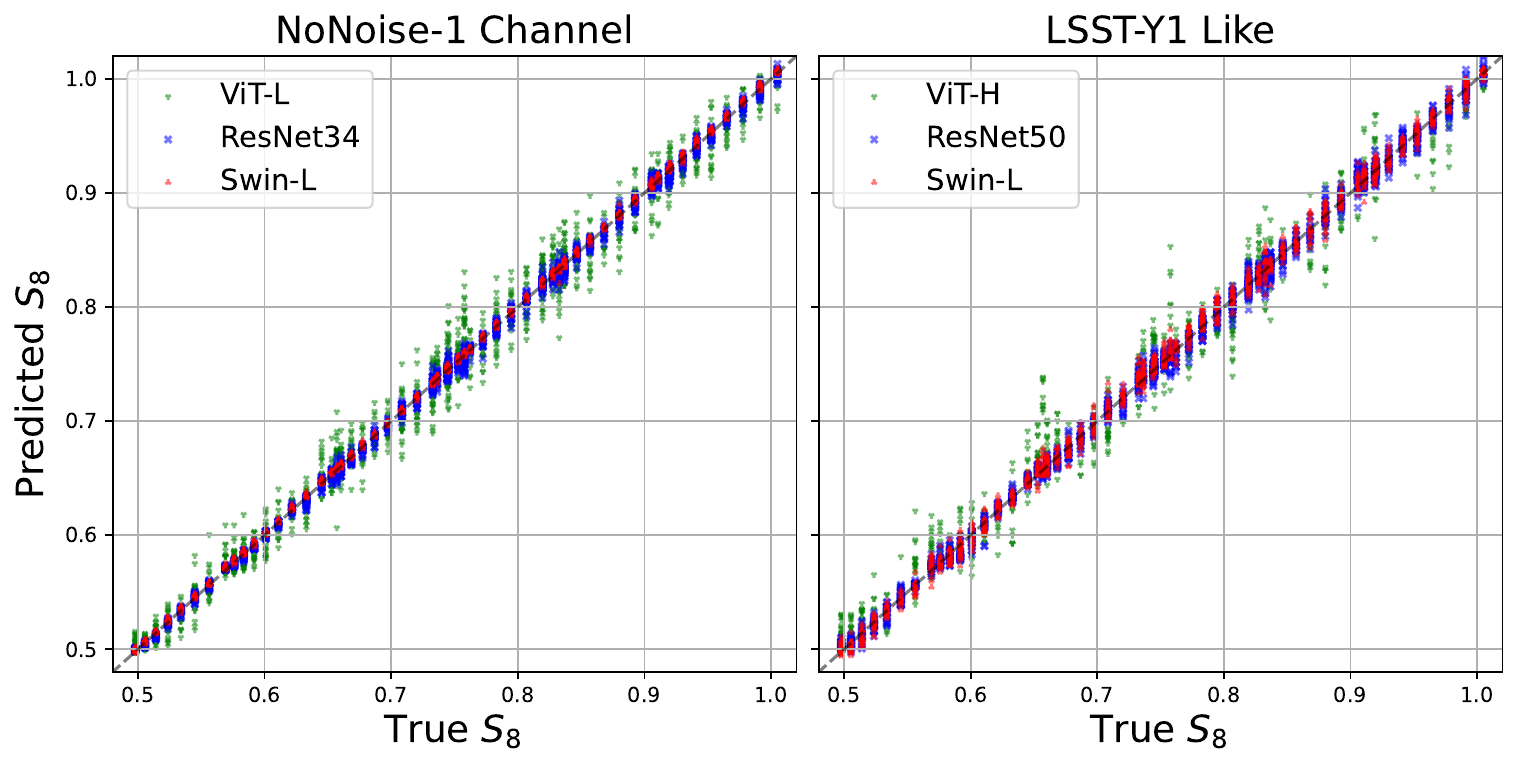}
    \includegraphics[width=0.49\linewidth]{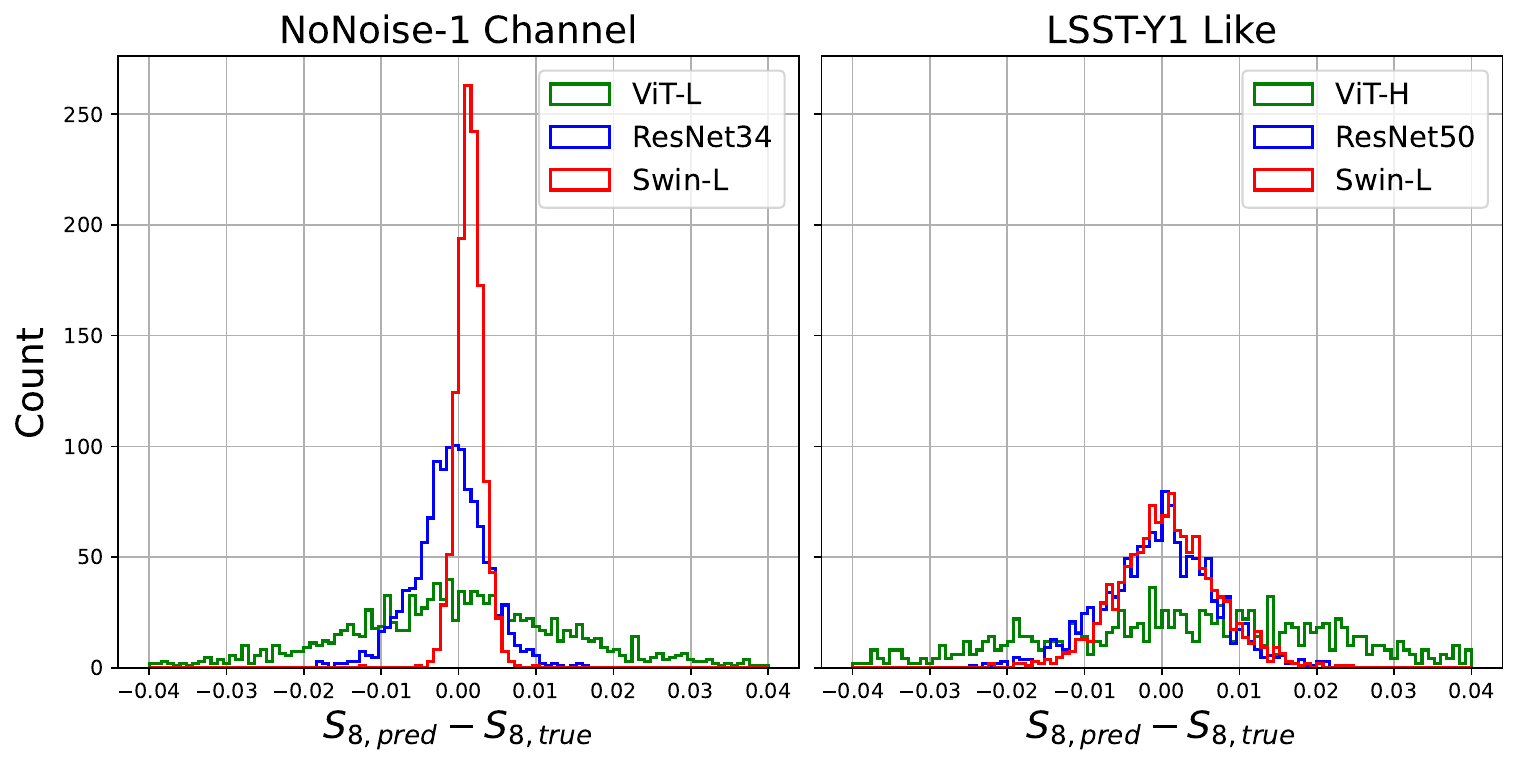}
    \caption{The vision model prediction of the parameter $S_8=\sigma_8\sqrt{\Omega_\mathrm{m}/0.3}$. The first two plots show the prediction vs the true value, and the last two plots show the histogram of the residuals. The statistical measures are reported as root mean-square error and $R^2$ values, and we summarize them in Appendix~\ref{appendix:model_exploration}. Note that \textit{NoNoise} is an exploratory case with only one channel, and \textit{LSST-Y1} Like contains the maps with 4 channels, as detailed in Section \ref{sec:methodology}.}
    \label{fig:best_models}
\end{figure}

\begin{figure}
    \centering
    \includegraphics[width=0.49\linewidth]{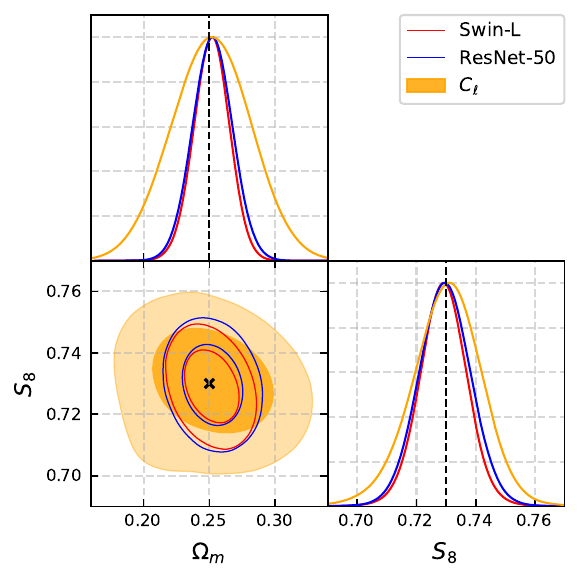}
    \includegraphics[width=0.49\linewidth]{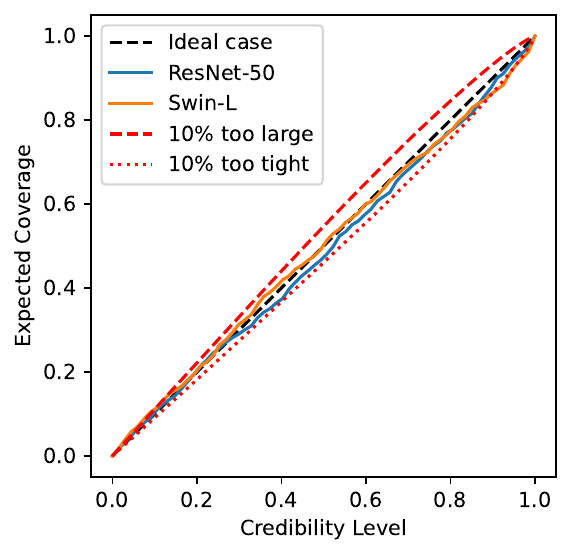}
    \caption{\textit{Left}: SBI posteriors obtained from our NDE setup detailed in Section~\ref{sec:methodology}, for a CNN (blue) and a Swin-Transformer (red) for cosmological weak lensing $\kappa$ fields. \textit{Right}: Results from TARP \cite{lemos2023samplingbasedaccuracytestingposterior}, which compares empirical coverage against credibility level to gauge posterior performance. Posteriors obtained on the test set from our NDE setup yield curves closely aligned with the identity line, indicating that neither of the posteriors is under- or over-constrained.}
    \label{fig:posteriors}
\end{figure}
\section{Attention-Based Vision Models}\label{sec:models}

Transformers, first introduced in \cite{vaswani2017attention}, have achieved remarkable success in natural language processing (NLP). They process sequential inputs—such as words in a sentence—using self-attention, which models contextual relationships between tokens. Each token is embedded and transformed via three learned matrices: the query, key, and value. Together, these form an attention head, which enables the model to capture pairwise dependencies across the input sequence. Similarly to how CNNs can learn different types of features using multiple filters (i.e., learned weight kernels), multiple learnable attention heads can capture different relations between the input tokens. Positional encodings are added to preserve order information, which is otherwise absent in the architecture.

To adapt transformers to computer vision, images are divided into fixed-size patches. Each patch is flattened and linearly projected into an embedding space. These patch embeddings, along with positional encodings, are passed through a stack of transformer encoder blocks. Each block consists of multi-head self-attention and feedforward (MLP) layers, along with residual connections and normalization layers. For full architectural details, we refer the reader to \cite{dosovitskiy_image_2021, vaswani2017attention}.

The Swin Transformer \cite{liu2021swintransformerhierarchicalvision} modifies the ViT structure to improve efficiency and scalability. Instead of applying global self-attention across all patches, Swin computes attention locally within non-overlapping windows. These windows shift between layers, allowing cross-window interactions while preserving locality. This design reduces the computational complexity of self-attention from quadratic to linear in image size, enabling Swin to scale to high-resolution inputs with fewer parameters than ViT.

While more efficient than ViT, Swin models still have significantly more parameters than typical convolutional architectures. They can also exhibit training instabilities when scaled up. SwinV2 \cite{liu2022swintransformerv2scaling} addresses this issue, introducing improved normalization and scaling techniques for stable training at larger model sizes.

Both ViT and Swin models have much higher parameter counts than standard CNNs. For instance, the smallest ViT model used here has approximately 86 million parameters, while ResNet-50 has about 24 million. This makes architecture tuning impractical in our setting, where simulation and computational resources are limited. Instead, we adopt standard configurations: three ViTs (ViT-B, ViT-L, ViT-H) from \cite{dosovitskiy_image_2021}, and four Swin models (Swin-T, Swin-S, Swin-B, Swin-L) from \cite{liu2021swintransformerhierarchicalvision}.
\section{Results and Conclusion}\label{sec:results}

\begin{figure}
    \centering
    \includegraphics[width=0.9\linewidth]{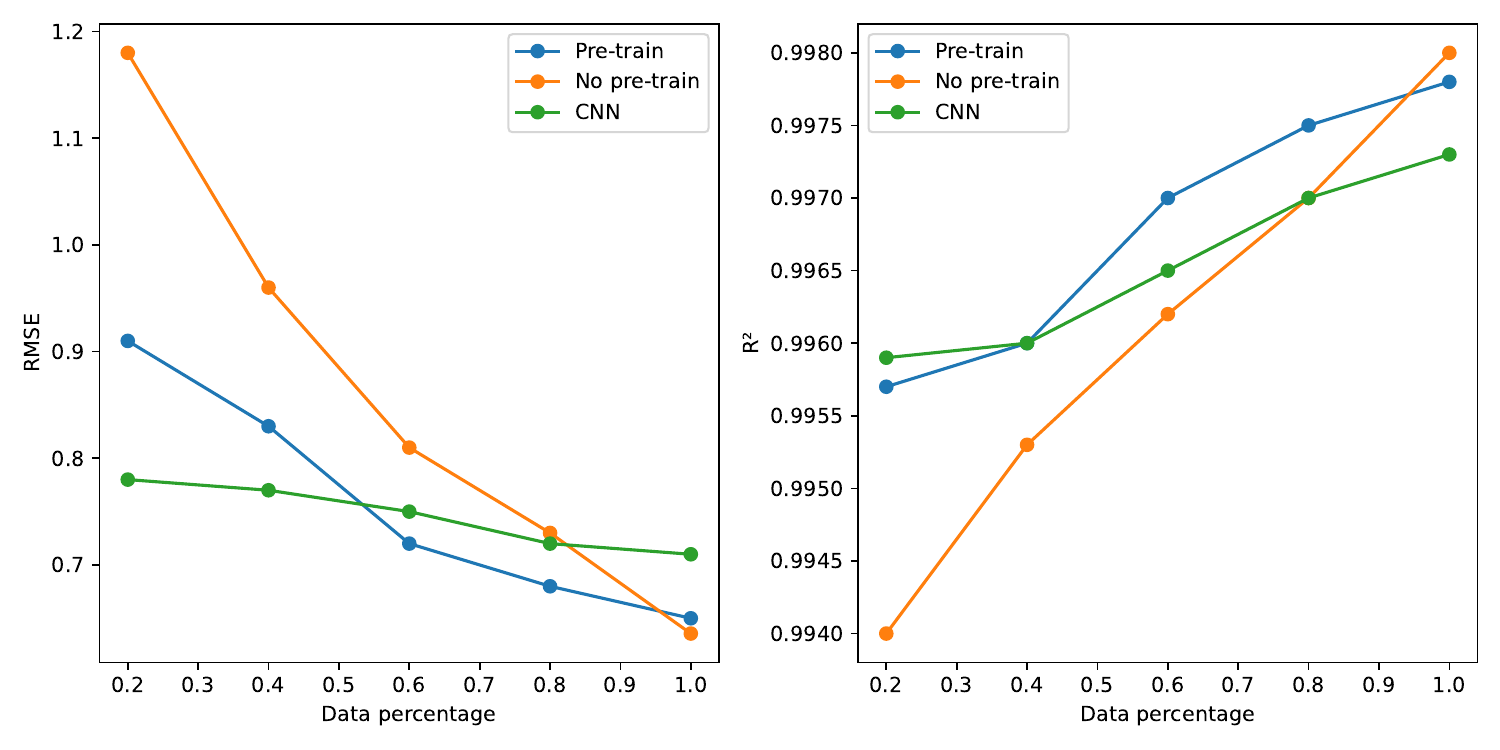}
    \caption{RMSE and $R^2$ for $S_8$ as a function of the training data fraction. Pre-training significantly improves the performance of attention-based models, especially in low-data regimes. In contrast, CNN baselines—shown without pre-training—exhibit stronger regularization and more stable performance.}
    \label{fig:pretraining}
\end{figure}

As shown in Figure~\ref{fig:best_models} and Figure~\ref{fig:posteriors}, the best attention-based model and the best convolution-based model achieve comparable constraining power. The full results are shown in Appendix~\ref{appendix:model_exploration}. Both of the models passed the coverage test that they are within 10\% of over-constraining. In the exploratory noiseless case, Swin Transformers initially appeared to outperform CNNs, likely due to their greater model flexibility. However, this advantage vanished once realistic shape noise was introduced. Vanilla Vision Transformers consistently underperformed relative to both CNNs and Swin models, possibly due to inefficient training on small datasets and difficulty capturing fine-scale features. 

We also find that pre-training significantly improves the performance of large models such as Swin Transformers, especially when training data are limited (Figure~\ref{fig:pretraining}). In contrast, pre-training has minimal effect on CNNs, consistent with previous findings \cite{Sharma:2024pth}. 

In conclusion, our experiments show that attention-based transformers do not outperform CNNs in constraining cosmological parameters from weak lensing convergence maps. With sufficient training data or pre-training, their performance can match that of CNNs. Although no gain in constraining power is observed, the interpretability features of transformer architectures remain a promising direction for future exploration.

\bibliography{ViT}

%%%%%%%%%%%%%%%%%%%%%%%%%%%%%%%%%%%%%%%%%%%%%%%%%%%%%%%%%%%%

\appendix

\section{Model Architectures}\label{appendix:model_complexity}

To adequately benchmark the performance of vision transformer models for weak lensing, we also explore a number of convolutional models. Convolutional Neural Networks (CNNs) are a standard deep learning tool for regressing parameters from cosmological and astrophysical multi-dimensional data \cite{zhong2024improvingconvolutionalneuralnetworks}, with built-in translation invariance and assumption of hierarchical local spatial information, due to the underlying convolution operation. As performance benchmarks, we use a simple baseline CNN with six convolutional layers feeding into a four-layer feedforward perceptron, with pooling and LeakyReLU activation functions. While we have not optimized the hyperparameters and architecture of this baseline CNN, this architecture has been shown to achieve competitive results for these types of tasks \citep{zhong2024improvingconvolutionalneuralnetworks}. 

In addition, we use residual convolutional neural networks (ResNets) introduced in \citet{he2015deep}. Contrary to naive expectations, increasing the depth of standard convolutional networks did not improve performance due to the degradation problem where an optimizer is unable to approach identity mapping; \citet{he2015deep} circumvented this by adding non-trainable residual connections that allow the model to \textit{skip} layers through connections of identity or projection mappings. Thus, ResNets are able to be deeper than conventional CNNs; \texttt{ResNet50} specifically has 50 layers structured in several \textit{bottleneck} blocks that are shortcutted by identity mappings. Beyond conventional computer vision tasks, residual convolutional networks have found contemporary applications in galaxy morphology classification \citep{barchi_machine_2020, Fielding2021, Ćiprijanović_2022}, wavefront sensing for image sharpening \citep{Andersen2020}, Ly-$\alpha$ emitter identification \citep{li_using_2019}, and Cosmic Microwave Background analysis \citep{Caldeira2019, he_analysis_2018}, making it an apt choice for our comparison standard. The full set of models, along with their respective parameter counts, are detailed in Table \ref{table:model_params}.

% \begin{figure}
%  \centering
%   \includegraphics[width=\columnwidth]{figs/baseline_cnn.drawio.pdf}
%   \label{fig:baseline_cnn}
% \end{figure}

\begin{table}[t]
  \centering
  \begin{tabular}{ll}
    \toprule
    Name     & Parameter Count     \\
    \midrule
    Baseline CNN & 500K \\
    ResNet-18     &  11M \\
    ResNet-34     & 21M \\
    ResNet-50     & 24M \\
    ResNet-101     & 44M \\
    ViT-B & 86M \\ 
    ViT-L & 307M \\ 
    ViT-H & 632M \\ 
    Swin-T & 29M \\
    Swin-S & 50M \\
    Swin-B & 88M \\
    Swin-L & 197M \\
    \bottomrule
  \end{tabular}
  \caption{Model names and number of parameters}
  \label{table:model_params}
\end{table}
\section{Detailed Model Performance}\label{appendix:model_exploration}
We test our models with three simulation cases. For the first case (\texttt{NoNoise--1 Channel}), we choose not to include observational shape noise or tomography. For the second (\texttt{LSST-Y1 1 Channel}), we include shape noise at the expected level of LSST-Y1, with effective source number density $n_{\text{eff}} \sim 12$. For both, we use a redshift kernel similar to the DES third tomography bin \cite{Myles_2021}. The final case (\texttt{LSST-Y1 Like}), we implement a full tomographic map-making procedure, generating 4-channel maps, each corresponding to the four DES tomographic bins. 

We test each model outlined in Table \ref{table:model_params} against each simulation case. For each configuration, we use an 80\%:10\%:10\% for our training, validation, and testing split. Each input is a $512\times512$ independent flattened patch of a convergence map, with either one or four channels. We train each network using the AdamW optimizer \cite{loshchilov2019decoupledweightdecayregularization}, with an early stopping patience of 30 epochs. A  L2 error metric (root-mean-squared, RMSE) is used, computed over the validation dataset. Batch size is 32, except for the much larger networks \texttt{ViT-L} and \texttt{ViT-H} (each $>300$M parameters), where we set to 24 and 12 respectively, due to GPU memory constraints (we use Nvidia A100s with 80 GBs of memory). We use the PyTorch ReduceLROnPlateau scheduler with a 10-epoch patience and reduction rate of 0.3. For all convolutional networks and vision transformers, the initial learning rate was set to $10^{-3}$ and $10^{-5}$ respectively. Other values were set to PyTorch defaults. After running the full sweep, the Swin-L model was observed to achieve the lowest RMSE value. See Table \ref{table:ml_performance} for the full results.

\begin{table}[t]
    \centering
    \resizebox{\textwidth}{!}{
    \begin{tabular}{||c||c c c c c c c c c c c c||}
        \hline
        Setting & B-CNN & RN-18 & RN-34 & RN-50 & RN-101 & ViT-B & ViT-L & ViT-H & Swin-T & Swin-S & Swin-B & Swin-L \\
        \hline\hline
        NoN (RMSE) & 0.6298 & 0.5239 & 0.4724 & 0.5084 & 0.5294 & 1.860 & 1.549 & 1.662 & 0.3724 & 0.3088 & 0.3209 & \textbf{0.2232} \\
        NoN ($R^2$) & 0.9981 & 0.9987 & 0.9989 & 0.9987 & 0.9986 & 0.9831 & 0.9883 & 0.9865 & 0.9993 & 0.9995 & 0.9994 & 0.9998 \\
        \hline
        T1 (RMSE) & 0.7436 & 0.7852 & 0.6170 & 0.6227 & 0.7613 & 1.725 & 1.621 & 1.859 & 0.7766 & 0.6759 & 0.6480 & \textbf{0.5492} \\
        T1 ($R^2$) & 0.9973 & 0.9970 & 0.9981 & 0.9981 & 0.9972 & 0.9855 & 0.9872 & 0.9832 & 0.9971 & 0.9978 & 0.9980 & 0.9985 \\
        \hline
        % T4 (RMSE) & 1.272 & 1.244 & 0.964 & 1.317 & 1.211 & 2.304 & 2.294 & 2.284 & 1.524 & 1.334 & 1.352 & 1.260 \\
        % T4 ($R^2$) & 0.991 & 0.992 & 0.995 & 0.991 & 0.992 & 0.972 & 0.972 & 0.973 & 0.988 & 0.991 & 0.990 & 0.992 \\
        % \hline 0.5740, 0.998
        T4 (RMSE) & - & 1.058 & 0.8986 & 0.7118 & 0.8734 & 2.005 & 2.105 & 1.988 & 1.061 & 1.021 & 0.8830 & \textbf{0.6357} \\
        T4 ($R^2$) & - & 0.995 & 0.996 & 0.998 & 0.996 & 0.980 & 0.978 & 0.981 & 0.995 & 0.995 & 0.996 & 0.998 \\
        \hline
    \end{tabular}
    }
    \caption{Performance comparison of different ML models across various settings.}
    \label{table:ml_performance}
\end{table}

\end{document}